# Tracing Methamphetamine abuse in under-treatment drivers: How biomechanical and oculomotor features help detect at-risk drivers?

**Hamed Salmanzadeh[1*], Alireza mortezapour[1,2], Iman Tahbazzadeh moghaddam[3], Farshid Ipackchi[1], Payam Abedinzadeh[1], Samira teimoori[3]**

[1]Faculty of Industrial Engineering, K. N. Toosi University of Technology, Tehran, Iran

[2]Department of Computer Science, University of Salerno, Fisciano, Italy

[3]Faculty of Mechanical Engineering, K. N. Toosi University of Technology, Tehran, Iran

[*]**: Corresponding author:** Hamed Salmanzadeh; h.salmanzadeh@kntu.ac.ir

**Abstract**

While the detrimental impacts of driving under the influence of stimulants such as methamphetamine are well-documented, the driving performance of individuals currently under-treatment has received considerably less attention. This study compared the behavior of individuals with a history of stimulant abuse (across two distinct treatment phases) with a control group of healthy drivers using a driving simulator. Oculomotor and biomechanical data were continuously collected via an eye-tracker and a Kinect sensor, respectively. These parameters were utilized to train a K-Nearest Neighbors (KNN) classification model designed to detect high-risk behavioral patterns in drivers undergoing methamphetamine rehabilitation. Through the evaluation of various feature combinations and neighborhood configurations, the optimized model successfully discriminated between normal drivers and those with a history of abuse with an accuracy of 90%. Detecting at-risk drivers through technologies embedded in Advanced Driver Assistance Systems (ADAS) by continuously monitoring physiological and behavioral parameters, facilitates a proactive safety strategy. Issuing real-time alerts to the driver, passengers, and external monitoring networks can ultimately mitigate the risk of traffic collisions.

## 1. Introduction

Methamphetamine is among the most widely used illicit stimulants worldwide, with its production, trafficking, and consumption increasing steadily over the past decade [1, 2]. Between 2010 and 2020, both the global consumption of methamphetamine and the number of individuals seeking treatment for dependence exhibited a continuous upward trajectory [3]. According to the United Nations Office on Drugs and Crime (UNODC), approximately 292 million individuals used illicit drugs in 2022 - a 20% increase from the previous decade - with amphetamines ranking as the third most consumed illicit substance worldwide [1, 4]. Consequently, the World Health Organization (WHO) has recognized driving under the influence of drugs (DUID) as a rapidly growing global road safety concern [2, 5]. The sheer scale of this issue is evident across the globe. In the United States, the country has entered the fourth wave of the opioid epidemic, accompanied by a surge in stimulant prevalence from 9.3% to 25.1% [4]. The National Survey on Drug Use and Health reported that 5.6% of drivers over the age of 16 drove under the influence of illicit drugs in the past year [4]. Similarly, the European DRUID project highlighted alarming prevalence rates of substance use among severely injured drivers, reaching 22.4% in Belgium, 19.2% in Finland, and 16.2% in Denmark [6]. A five-year study in Denmark (2015–2019) recorded a 68% increase in DUID cases, with central nervous system (CNS) stimulants comprising 46.8% of these instances, driven primarily by cocaine (23.8%) and amphetamines (22.9%) [7].

The impact of these substances on traffic safety is devastating, and methamphetamine is now the most common illicit drug detected in injured or fatally injured drivers [4, 8, 9]. In the U.S., data from 683,184 drivers admitted to trauma centers between 2017 and 2020 showed that 28.8% tested positive for at least one substance, a figure that rose to 32% by 2020 [4]. Among these positive cases, 16.3% involved amphetamines and 4.8% involved methamphetamine, with 36.3% of drivers exhibiting poly-drug use. Recent evidence further indicates that polysubstance-impaired driving is becoming increasingly prevalent among fatally injured drivers in the United States, with approximately one in five fatally injured drivers exhibiting combined alcohol and drug impairment [10]. In Spain (2021), autopsies of 812 fatally injured drivers revealed that 49.4% were positive for at least one substance [11]. Furthermore, in Australia, DUI accounts for approximately 25% of all traffic crashes and 50% of fatal crashes [8]. A specific study of hospitalized drivers in Melbourne found that 10.1% tested positive for methamphetamine - more than 3.5 times higher than opioids - accounting for 76.9% of all illicit drug-positive cases [8]. In China, out of 896 DUID convictions, 71.2% involved amphetamine-type stimulants (ATS), resulting in 769 crashes and 629 fatalities [12].

While certain demographics features such as being a truck drivers (typically aged 20 to 41), may use low-to-moderate doses of amphetamines to combat sleepiness and temporarily improve processing speed [1], these effects deteriorate drastically at higher doses or during complex tasks like complex driving [2]. Methamphetamine consumers frequently exhibit hazardous behaviors, including high-speed driving, frequent lane changing, lane departure, failure to stop, unsafe merging, and increased risk-taking [8, 10]. Beyond impaired vehicle control, substance use has also been associated with increased driving anger, aggression, and road rage, further compromising traffic safety [13]. Unlike alcohol, blood concentrations of methamphetamine do not reliably predict crash risk; however, the comparative risks are staggering. While a blood alcohol concentration (BAC) of 0.5 g/L increases crash risk by approximately 1.4 times, amphetamines escalate this risk by 5 to 30 times [8, 12]. According to data from Norway, amphetamines carry a relative crash mortality risk of 5.70, the highest among evaluated substances, far exceeding cannabis (1.39) and cocaine (2.91) [14]. Overall, methamphetamine-impaired drivers are 5 times more likely to die in a crash, over 6 times more likely to be injured, and up to 23.5 times more likely to be deemed culpable in collisions [2, 15].

Beyond acute intoxication, chronic dependence and the subsequent withdrawal or treatment phases present critical, yet often overlooked, safety challenges [11]. Problematic drug use is the strongest predictor of DUID, and studies indicate that drug offenders have a rearrest rate of 18% to 29%, compared to just 6% to 12% for alcohol offenders, proving that this is a persistent behavioral issue [11, 16]. A review of studies confirmed that chronic methamphetamine use impairs attention, executive function, memory, reaction time, processing speed, and inhibitory control, while increasing psychosis, anxiety, paranoia, and hallucinations [17, 18]. A driving simulator study further reported that chronic users exhibit poorer speed control, worse lane positioning, and inferior gap acceptance [19]. These cognitive deficits frequently manifest as aggressive, reckless, anxious, and dissociative driving [8, 15]. Despite this, the overwhelming majority of previous research has focused exclusively on acute intoxication, leaving the role of the withdrawal phase and the behavior of drivers actively undergoing treatment critically under-investigated.

Identifying impairment before a crash occurs has become a critical challenge for next-generation intelligent transportation systems [8, 20]. Because problematic drug dependence is a deeply rooted issue, traditional law enforcement measures such as increased police presence or external biochemical testing (e.g., blood or breath analyses) were insufficient [21, 22]. For instance, the actual probability of arresting a drug-impaired driver is extremely low, estimated at fewer than 10 cases per one million kilometers driven in Norway [23]. These limitations highlight the need to complement conventional roadside testing with proactive in-vehicle monitoring technologies. Monitoring physical and behavioral changes, such as utilizing eye-tracking to reveal how substances impact driver attention and awareness, offers a highly proactive approach [20]. Despite the increasing application of eye tracking and driver monitoring systems in detecting distraction, fatigue, and drowsiness [24, 25], little is known about whether oculomotor behavior can objectively characterize drivers undergoing treatment for stimulant abuse.

If these high-risk behaviors are detected in real-time, the data can be seamlessly embedded into ADAS [26]. Through continuous physiological and behavioral monitoring, ADAS can establish a proactive safety by issuing immediate warnings to the driver, passengers, and external monitoring systems, thereby significantly reducing the likelihood of collisions. Given that at-risk drivers undergoing rehabilitation for stimulant abuse have received far less attention than active consumers, the primary objective of this study was to identify and classify the distinct oculomotor and biomechanical parameters of drivers under treatment for methamphetamine abuse.

## 2. Materials and Methods

### 2.1. Participants

The study population comprised individuals with substance use disorders (SUD) referred to the health center of the Congress 60 Human Revivification Society. Within this organization, patients undergoing treatment are referred to as "passengers". The treatment protocol consists of a gradual tapering method utilizing Opium Tincture (OT) over a 10- to 12-month period, administered in accordance with Congress 60 guidelines. Based on pharmacological standards, OT is a bitter hydroalcoholic solution generally composed of 10% opium (equivalent to 1% morphine), 20% alcohol, and 70% distilled water, along with zinc and preservatives.

For the experimental group, 12 male volunteers currently undergoing treatment at Congress 60 were selected. Additionally, 12 healthy males were recruited to serve as the control group. The control group was matched with the experimental group in terms of age and gender in order to control for the confounding effects of these demographic variables on the patterns of amphetamine use and the cognitive abilities required for driving tasks [15, 17].

The inclusion criteria for all participants required the absence of specific infectious diseases, no concurrent pharmacological treatments, and a stable family environment. Participants were required to be free of significant familial and social tensions, as such stressors could potentially confound the study's outcomes and lead to exclusion from the research [8].

The demographic and clinical characteristics of the participants are presented in Table 1. In this table, the experimental group (passengers) is further divided into two subgroups based on their treatment duration: individuals who have undergone less than 60 days of treatment, and those who have completed at least 60 days since the onset of their treatment.

Participation in this study was entirely voluntary without any monetary compensation. The research protocol was officially approved by the Ethics Committee of BLINDED University of Technology. Prior to enrollment, all participants were fully informed about the study's objectives and procedures, and they provided their written informed consent by signing the official consent forms.

**Table 1**. The participants' information

| **Parameter** | **Control Group (n=12)** | **Experimental (< 60 days) (n=6)** | **Experimental (≥ 60 days) (n=6)** |
|---|---|---|---|
| **Mean (Range)Age** | 30.4 (19 - 52) | 36.5 (23 - 50) | 36.0 (23 - 57) |
| **Days Since Treatment Onset** Mean (Range) | - | 34.7 (20 - 52) | 84.3 (61 - 135) |
| **OT Frequency (doses/day)** Mode (Range) | - | 3 (3 - 5) | 3 (1 - 3) |
| **OT Dosage per Admin. (cc)** Mean (Range) | - | 1.5 (1.0 - 2.0) | 2.3 (0.2 - 3.5) |
| **Predominant Substances Used** | - | Methamphetamine, Opium | Methamphetamine, Opium, Heroin |

## 2.2.Driving simulator and the simulation environment

The experiment utilized a low-fidelity driving simulator (Figure 1). The physical setup comprised a single large monitor to display the virtual road environment, paired with commercial-grade steering wheel and pedal controllers for manual driving.

The experimental protocol commenced with a familiarization phase, allowing participants to acclimate to the environment and vehicle dynamics; data from this practice session were excluded from the final analysis. For the main trials, drivers were instructed to maintain a "normal driving" behavior, remain in the center lane, and regulate their speed between 60 and 90 km/h. The simulated track included exactly 14 obstacles (e.g., pedestrians and domestic animals). Drivers were explicitly required to avoid collisions solely through appropriate steering maneuvers by executing safe lane changes without applying the brakes.

To effectively differentiate the cognitive workload and arousal levels between the experimental and control groups, and to enrich the feature space for the KNN classification algorithm, auditory stimuli were introduced as environmental stressors. Accordingly, participants completed three distinct scenarios: Scenario 1 featuring upbeat music, Scenario 2 featuring calm music, and Scenario 3 serving as a baseline condition with no music. This manipulation of auditory workload helps to unmask subtle neuro-motor variations between the groups. To prevent confounding variables while simultaneously eliminating spatial memory effects, the fundamental nature and total number of obstacles (14) remained strictly constant across all conditions; however, the precise spatial layout of the obstacles was varied in each scenario.

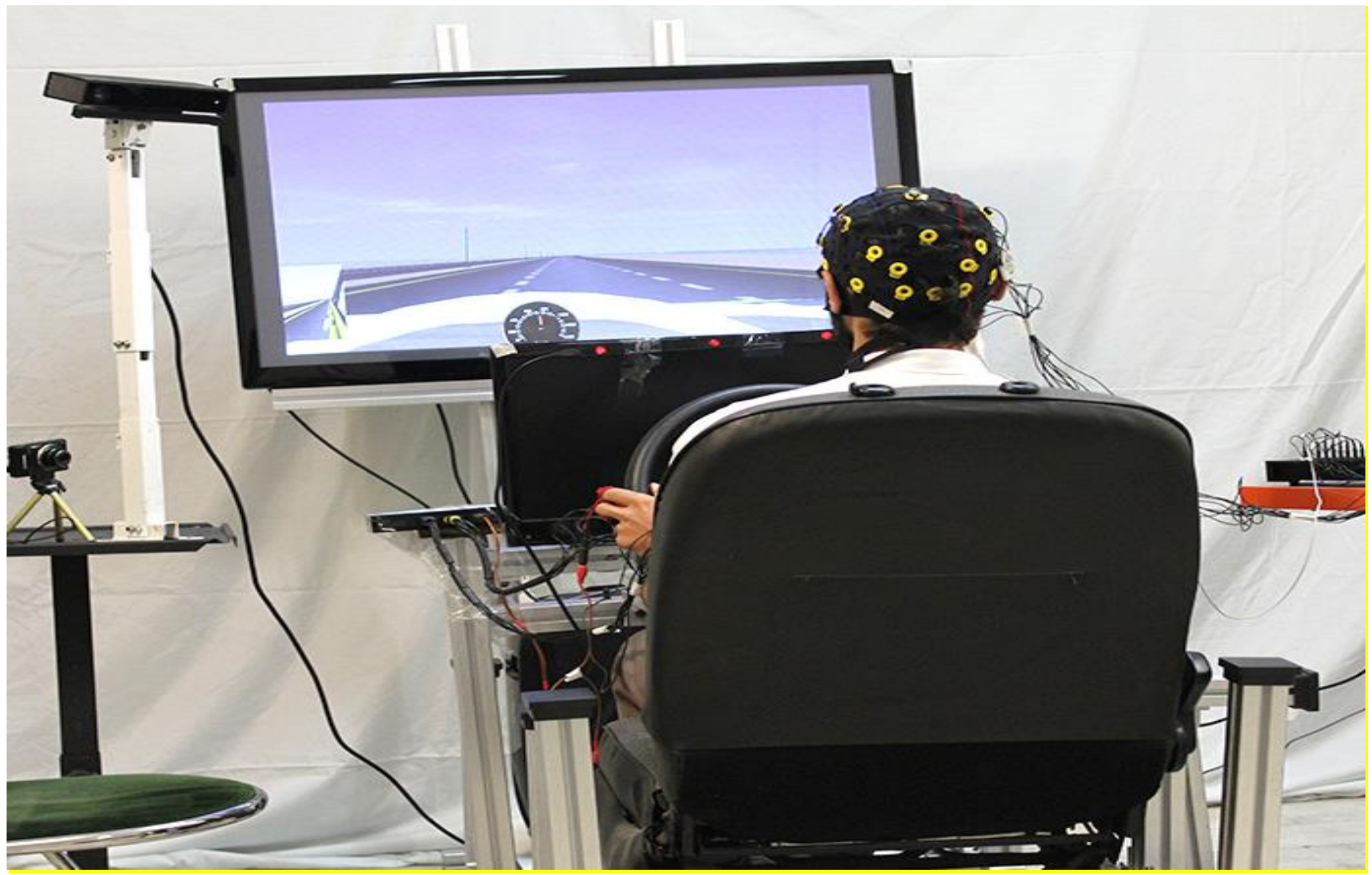

**Figure1** . Driving simulator and the external environment

### 2.3. Eye tracker

To monitor the drivers' visual behavior, an eye-tracking system was employed (Figure 2). The primary function of this apparatus is to determine the user's gaze point, effectively identifying the exact locus of their visual attention at any given moment within the simulated environment. The raw data extracted from the eye tracker are structured as two-dimensional arrays. The first and second dimensions of these arrays represent the horizontal and vertical coordinates of the gaze point on the display screen in pixels, which can be mathematically converted into physical millimeter units for precise spatial analysis.

These spatio-temporal data provide two critical streams of information for performance evaluation:

*Visual Attention Allocation:* It quantifies the extent to which the driver focuses on specific visual elements within the environment, such as rearview mirrors.

*Oculomotor Dynamics:* It reveals the temporal fluctuations and movement patterns of the eyeball throughout the duration of the experiment.

In the current study, a SteelSeries Sentry Eye Tracker (powered by Tobii technology) was utilized. As depicted in Figure 1, the device was mounted directly in front of the driver at the base of the monitor, capturing oculomotor data at a sampling frequency of 30 Hz.

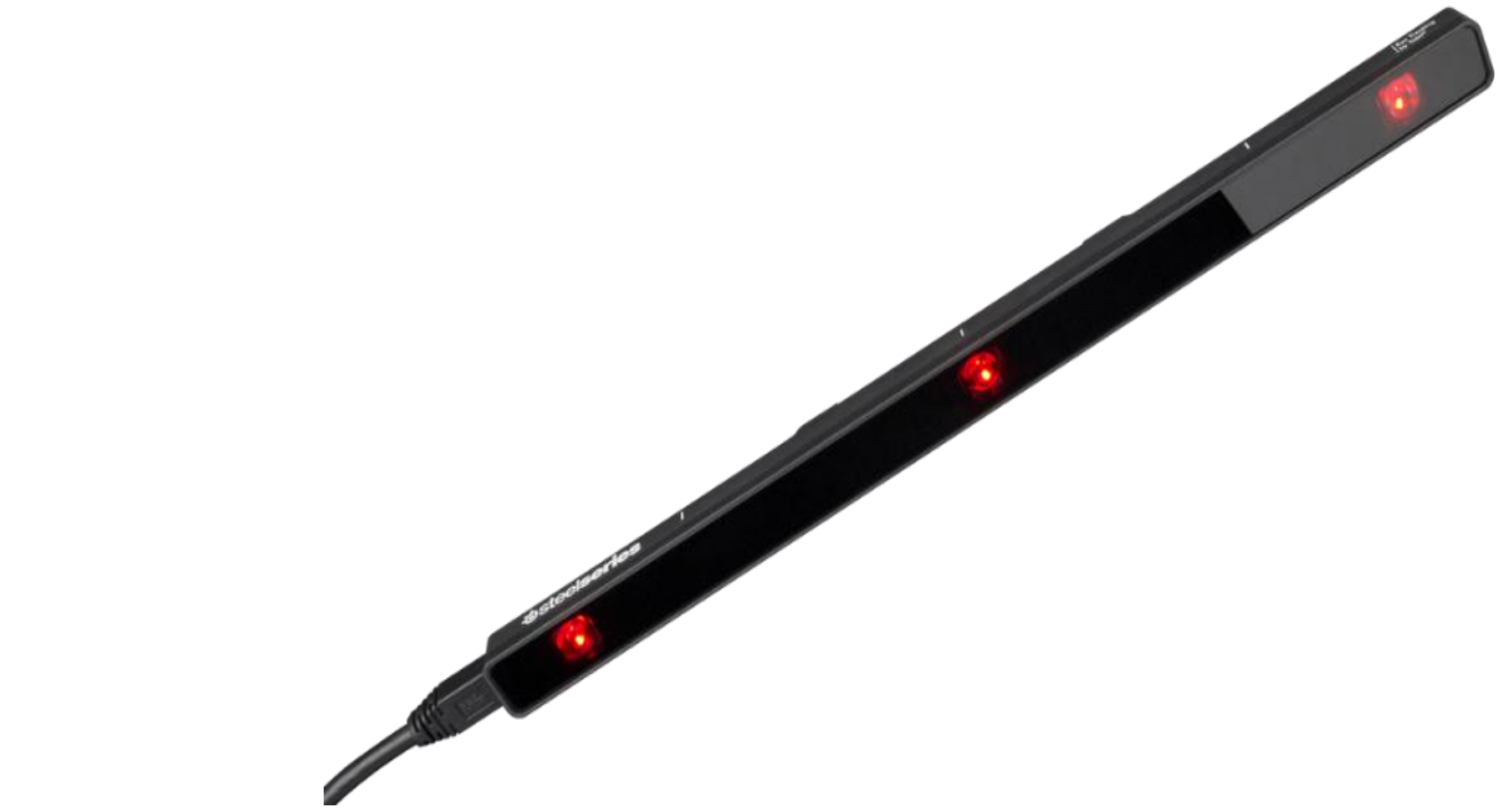

**Figure2 .** The eye tracker

### 2.4. Kinect Sensor

To record the driver's postural dynamics and biomechanical movements, a Kinect sensor was employed. The Kinect functions as an advanced, markerless motion-capture and voice-recognition system. It’s hardware architecture incorporates an RGB camera alongside an infrared (IR) depth sensor, which operate in tandem to capture high-definition imagery and generate accurate three-dimensional depth maps. While the device is traditionally designed to facilitate controller-free human-computer interaction and voice command processing [27], its primary application in the current study was to construct a real-time skeletal model of the driver and track their dynamic body movements. Figure 3 illustrates the extracted skeletal framework of a driver while seated in the simulator. The kinematic data and postural adjustments of the participants were continuously recorded by the sensor at a uniform sampling frequency of 30 Hz.

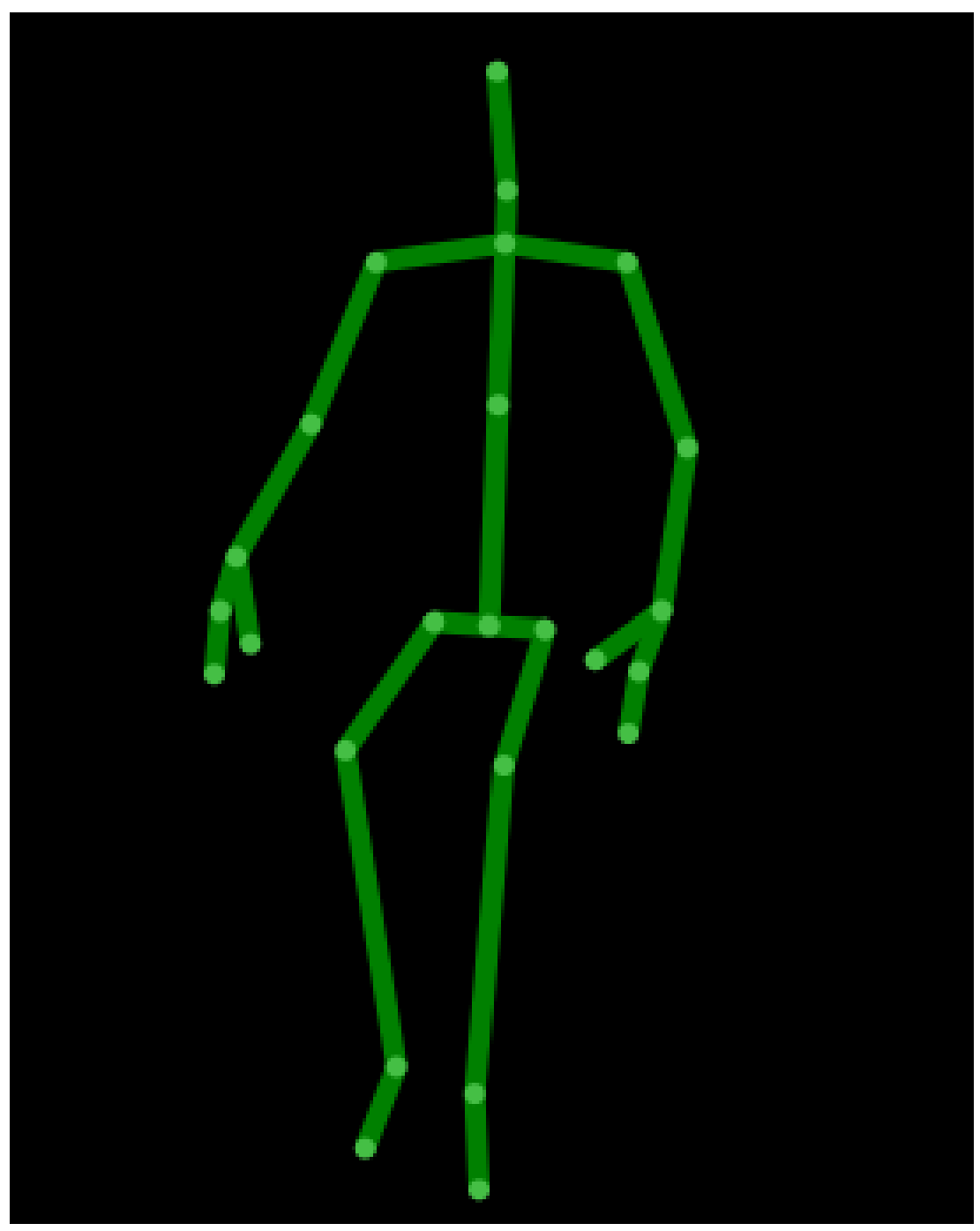

**Figure3 .** A sample of the extracted data from Kinect

## 2.5.Data Preprocessing and Feature Extraction

To develop a robust classification model, the raw data acquired from the Kinect and eye-tracking sensors were preprocessed and analyzed to extract discriminative features. Kinect Data Processing: The 3D spatial coordinates of the driver's left elbow joint were continuously recorded during the driving scenarios. Given the inherent noise in the raw Kinect data, a filtering pipeline was implemented prior to feature extraction. Initially, outlier data points were identified and removed using a statistical threshold defined in Equation 1, where mu and sigma represent the mean and standard deviation of the data array, respectively.

$$if\ (q_i < \mu - 3\sigma)\ or\ (q_i > \mu + 3\sigma)\ then\ q_i\ is\ outlier$$

**Equation 1.** Outlier detection rule

Following outlier removal, a moving average filter with a window size of 4 was applied (Equation 2) to smooth discontinuities and eliminate signal spikes. Figure 4 illustrates a comparison between the raw sensor readings and the filtered trajectory of the elbow joint along the x-axis.

$$y(t) = \frac{1}{4}\sum_{k=0}^{3} x(t-k)$$

**Equation 2.** Moving Average Filter

Subsequently, the cumulative displacement of the elbow joint was calculated using the filtered data according to Equation 3, where Delta t is the sampling time step of the Kinect sensor.

$$d(t_i) = \sqrt{\left(x(t_i + dt) - x(t_i)\right)^2 + \left(y(t_i + dt) - y(t_i)\right)^2 + \left(z(t_i + dt) - z(t_i)\right)^2}$$

**Equation 3.** How to calculate the displacement value

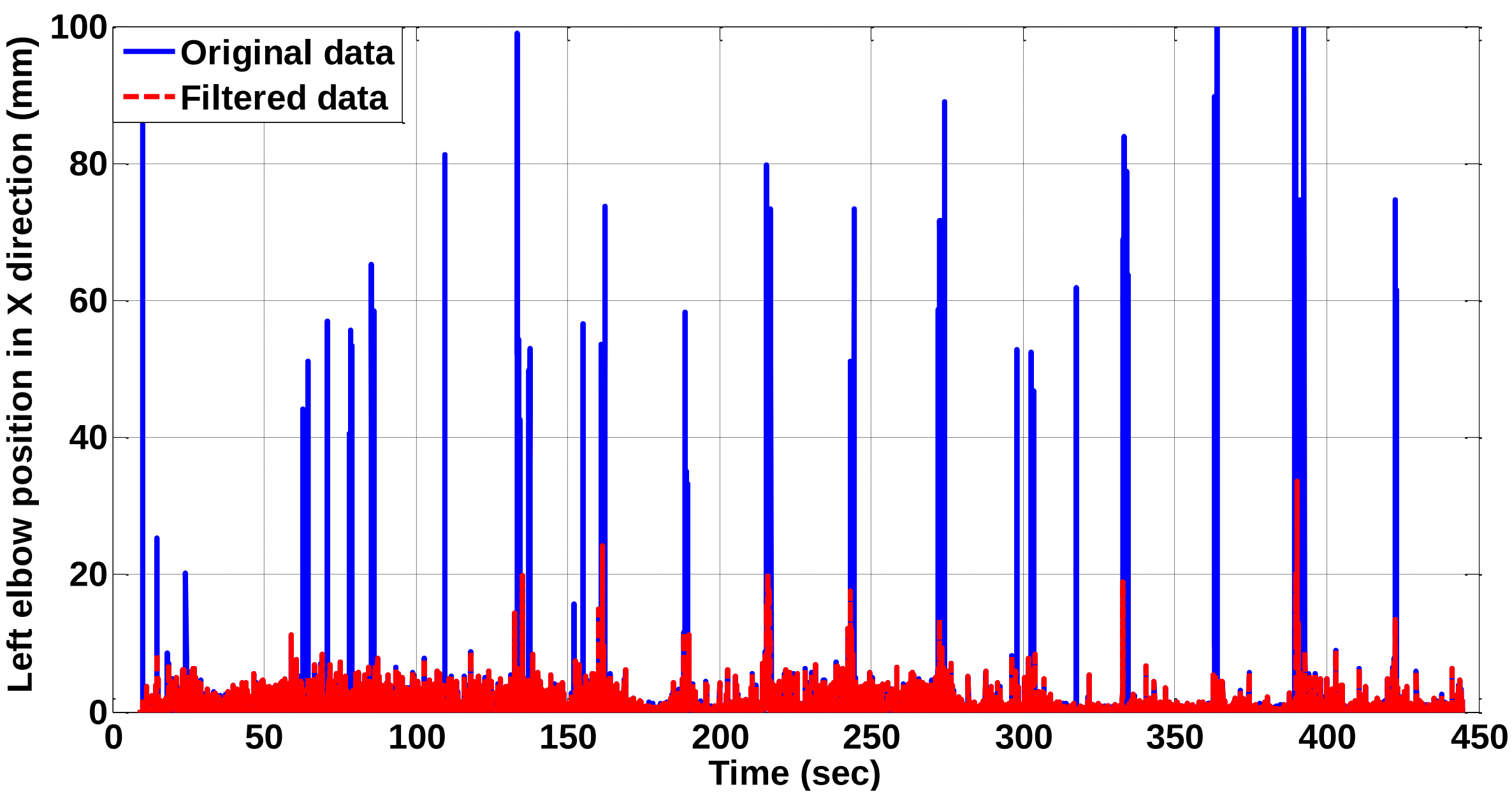

**Figure 4.** Comparison of sensor-read and filtered data of elbow joint position in the x-direction

The drivers' visual behavior was quantified using two primary metrics derived from the 2D gaze coordinates: pupil angular velocity and visual fixation. The angular velocity was calculated as the angular distance (in degrees) traversed by the pupil per unit of time. Visual fixation was defined by the duration the driver's gaze remained within specific radial thresholds (ranging from 3 to 8 cm on the screen) for time thresholds spanning 1 to 10 seconds.

A KNN classification algorithm was developed to categorize the participants based on the extracted behavioral and physiological features. Four key features were selected for model training, as detailed in Table 2.

**Table 2.** Number of features used in model design

| Feature No. | Feature Description |
|---|---|
| **1** | Percentage of the total journey duration during which the angular velocity of the pupil movement exceeded 10°/s. |
| **2** | Percentage of the total journey duration during which the eyes met the gaze criterion using a gaze duration threshold of 6 seconds and a gaze radius threshold of 7 cm (used for comparison between Groups 1 and 2). |
| **3** | Percentage of the total journey duration during which the eyes met the gaze criterion using a gaze duration threshold of 4 seconds and a gaze radius threshold of 8 cm (used for comparison between Groups 1 and 3). |
| **4** | Mean displacement of the left elbow joint. |

Prior to model training, all feature values were normalized to a scale of [0, 1]. The dataset was partitioned into a training set (70%) and a testing set (30%). Due to the limited sample size, the validation process was repeated for 100 iterations with randomized data shuffling to ensure that instances from all participant groups were adequately represented in both the training and testing phases.

The performance of the classification models was evaluated under various feature combinations and values of K (number of neighbors). The classification accuracy was determined using Equation 4.

$$Accuracy = \frac{1}{N}\sum\left(\frac{\sum_{k\in X_C} X_k}{\sum_{j\in X_T} X_j} \times 100\right)$$

**Equation 4.** How to calculate the accuracy of models

For the optimal model, a comprehensive evaluation was conducted using a Confusion Matrix. The model's predictive capabilities were further quantified using Sensitivity (True Positive Rate), Specificity (True Negative Rate), Precision (Positive Predictive Value), False Positive Rate (FPR), and overall Accuracy (Equations 5-9).

$$TPR = \frac{TP}{TP + FN}$$

**Equations 5.** How to calculate the sensitivity rate of a classification model

$$TNR = \frac{TN}{TN + FP}$$

**Equations 6.** How to calculate the recognition rate of a classification model

$$PPV = \frac{TP}{TP + FP}$$

**Equations 7.** How to calculate the accuracy rate of a classification model

$$FPR = 1 - TNR$$

**Equations 8.** How to calculate the false positive group detection rate of a classification model

$$ACC = \frac{TP + TN}{TP + FP + TN + FN}$$

**Equations 9.** How to calculate the accuracy rate of a classification model

## 3. Results

### 3.1. Biomechanical Features (Elbow Displacement):

The mean displacement of the left elbow joint across the participant groups is depicted in Figure 5 and summarized in Table 3. The results indicate that the combined mean displacement for the experimental groups (Groups 2 and 3) is 2.19 times greater than that of the control group (Group 1). A one-way ANOVA with a significance level of alpha = 0.05 was conducted. The p-value for comparing all three groups was 0.025, and the p-value for comparing Group 1 against the combined Groups 2 and 3 was 0.026. Therefore, the null hypothesis was rejected in both tests. The box plot in Figure 6 further corroborates the significant difference in elbow joint movement between healthy drivers and recovering patients, establishing this metric as a highly discriminative classification feature.

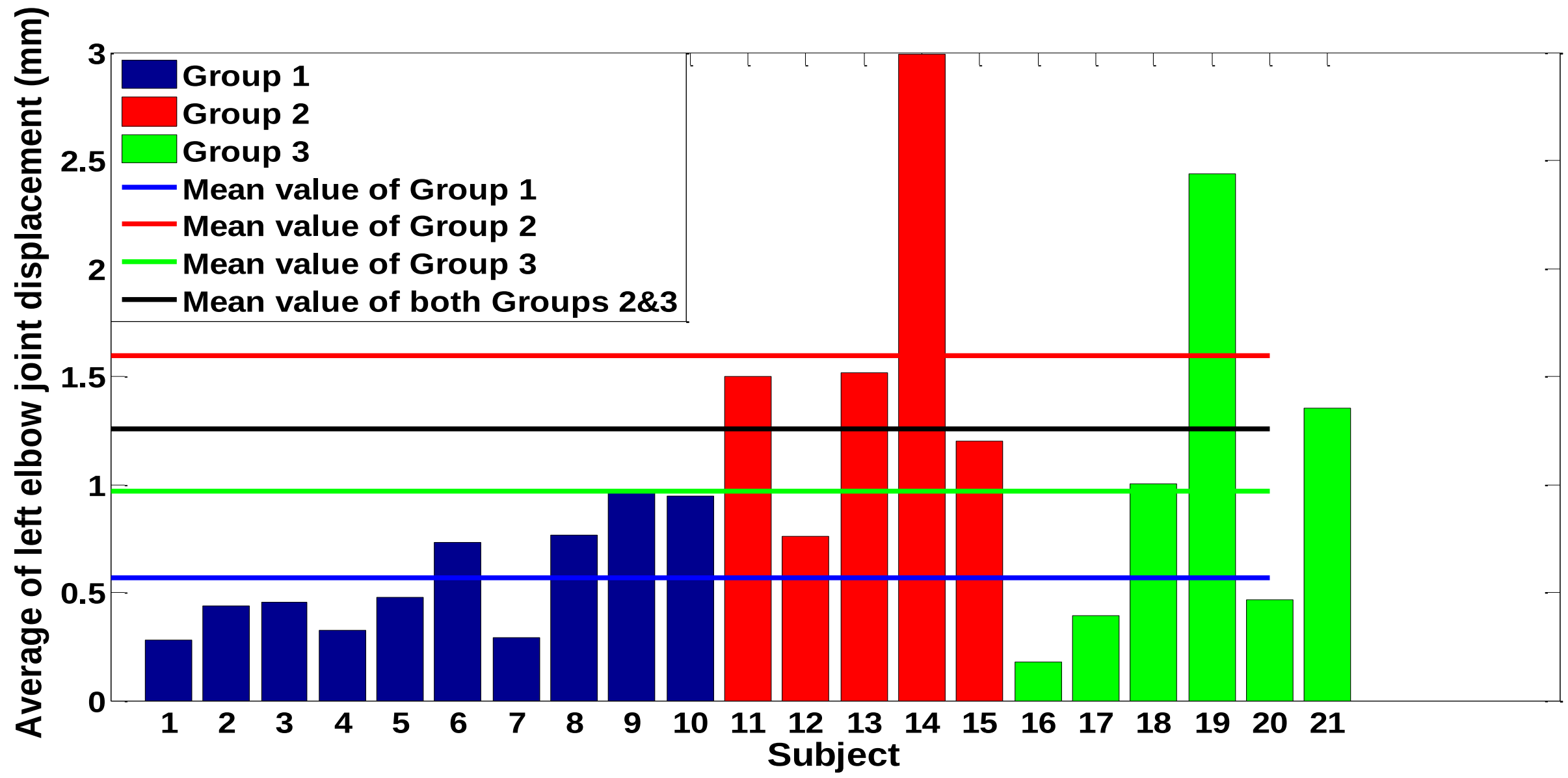

**Figure5 .**The average value of the left elbow joint displacement for individuals in all three groups

**Table3 .**Mean Left Elbow Joint Displacement for Participants in the Four Study Groups

| Group | Mean Left Elbow Joint Displacement (mm) |
|---|---|
| **1** | 0.57 |
| **2** | 1.60 |
| **3** | 0.97 |
| **2, and 3** | 1.25 |

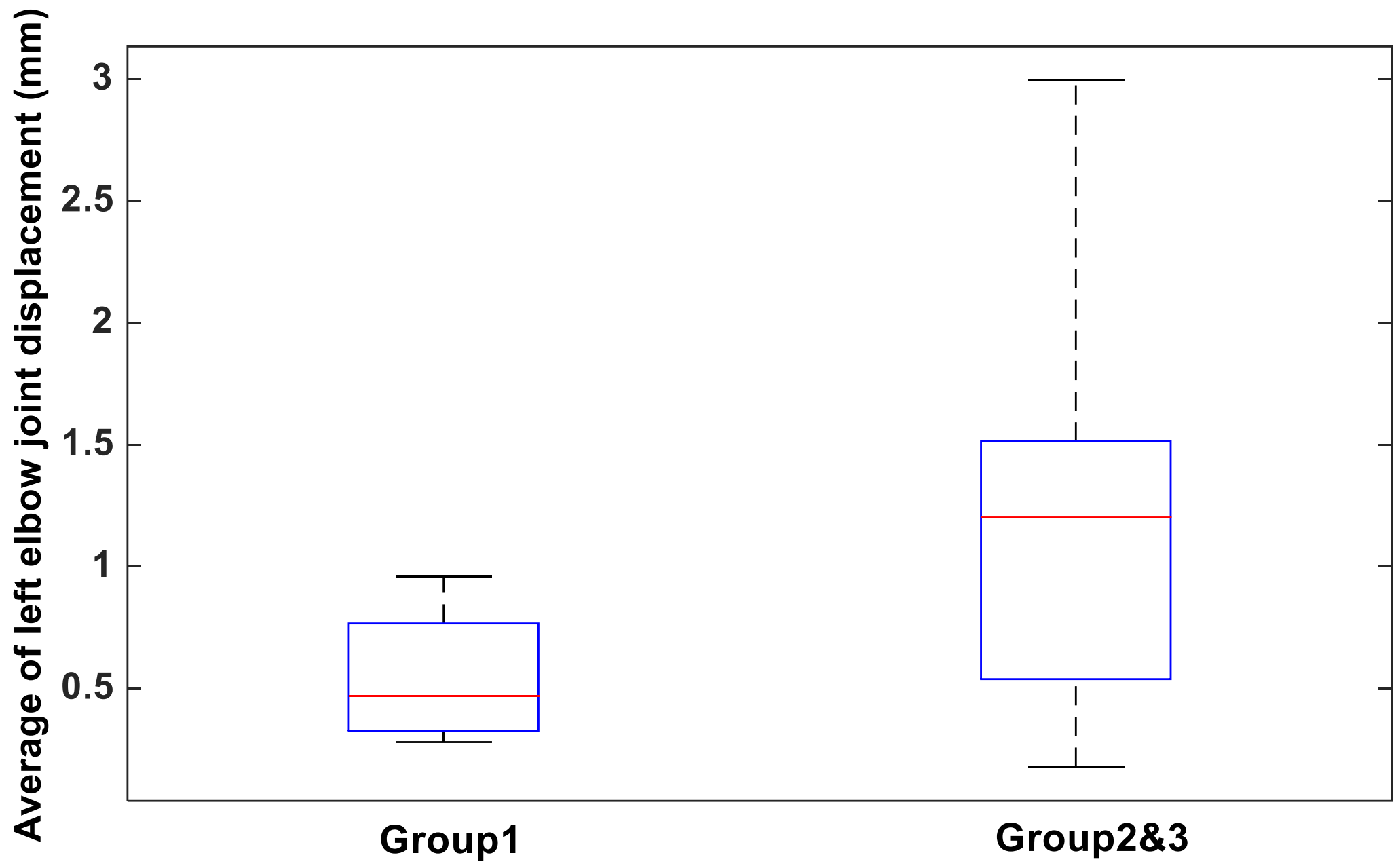

**Figure 6 .** Box diagram of the average displacement of the left elbow joint for people in group 1 and people in groups 2 and 3

### 3.2. Oculomotor Features (Angular Velocity and Fixation):

The analysis of pupil angular velocity across 9 different thresholds (5 to 45 deg/s) is presented in Table 4 and Figure 7. The data reveal that Group 2 (patients in the early stages of treatment, <2 months) exhibited significantly faster eye movements compared to the control group across all thresholds. Interestingly, Group 3 (long-term treatment) demonstrated slower eye movements than the healthy control group, suggesting a potential long-term physiological adaptation to the OT treatment. Figure 8 highlights the maximum percentage differences between the groups, occurring at the 10 deg/s threshold.

To further illustrate this contrast, Figure 9 presents the specific percentage of time each group's angular velocity exceeded the 10 deg/s threshold, highlighting the relatively homogenous behavior of Group 2 participants.

**Table 4.** Mean Percentage of Journey Duration During Which the Angular Velocity of Pupil Movement Exceeded Different Threshold Values Across the Study Groups

| ***Group*** | **>5°/s** | **>10°/s** | **>15°/s** | **>20°/s** | **>25°/s** | **>30°/s** | **>35°/s** | **>40°/s** | **>45°/s** |
|---|---|---|---|---|---|---|---|---|---|
| ***1*** | 66.9 | 45.1 | 31.5 | 23.2 | 17.7 | 13.8 | 11.1 | 9.2 | 7.8 |
| ***2*** | 75.3 | 54.2 | 39.8 | 29.6 | 22.6 | 17.7 | 14.1 | 11.5 | 9.6 |
| ***3*** | 60.3 | 38.3 | 26.4 | 19.4 | 14.8 | 11.8 | 9.7 | 8.1 | 6.9 |
| ***2, and 3*** | 67.8 | 46.2 | 33.1 | 24.5 | 18.7 | 14.7 | 11.9 | 9.8 | 8.3 |

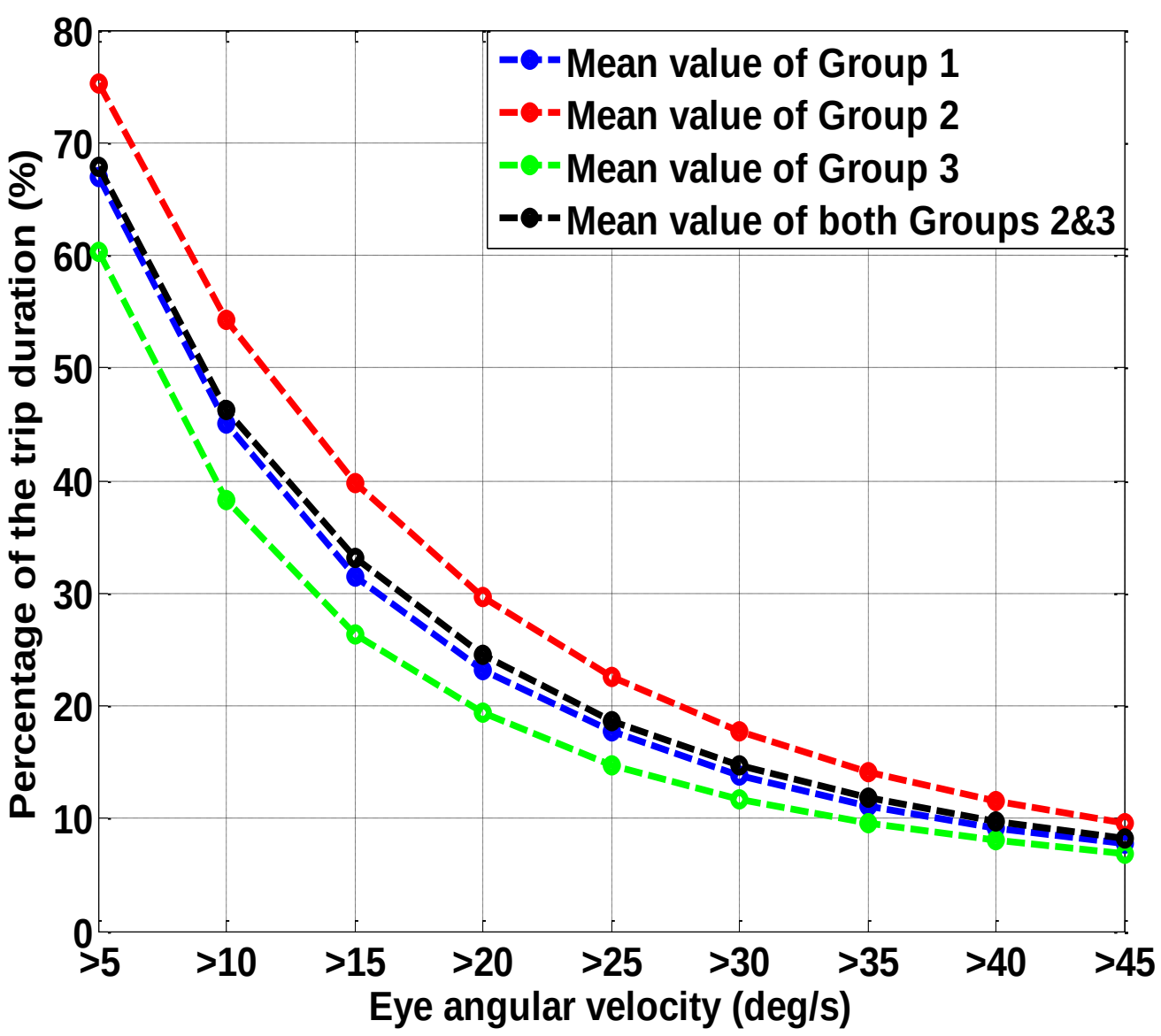

**Figure 7.**Average index of percentage of travel time for different thresholds of pupil angular velocity for individuals in three groups

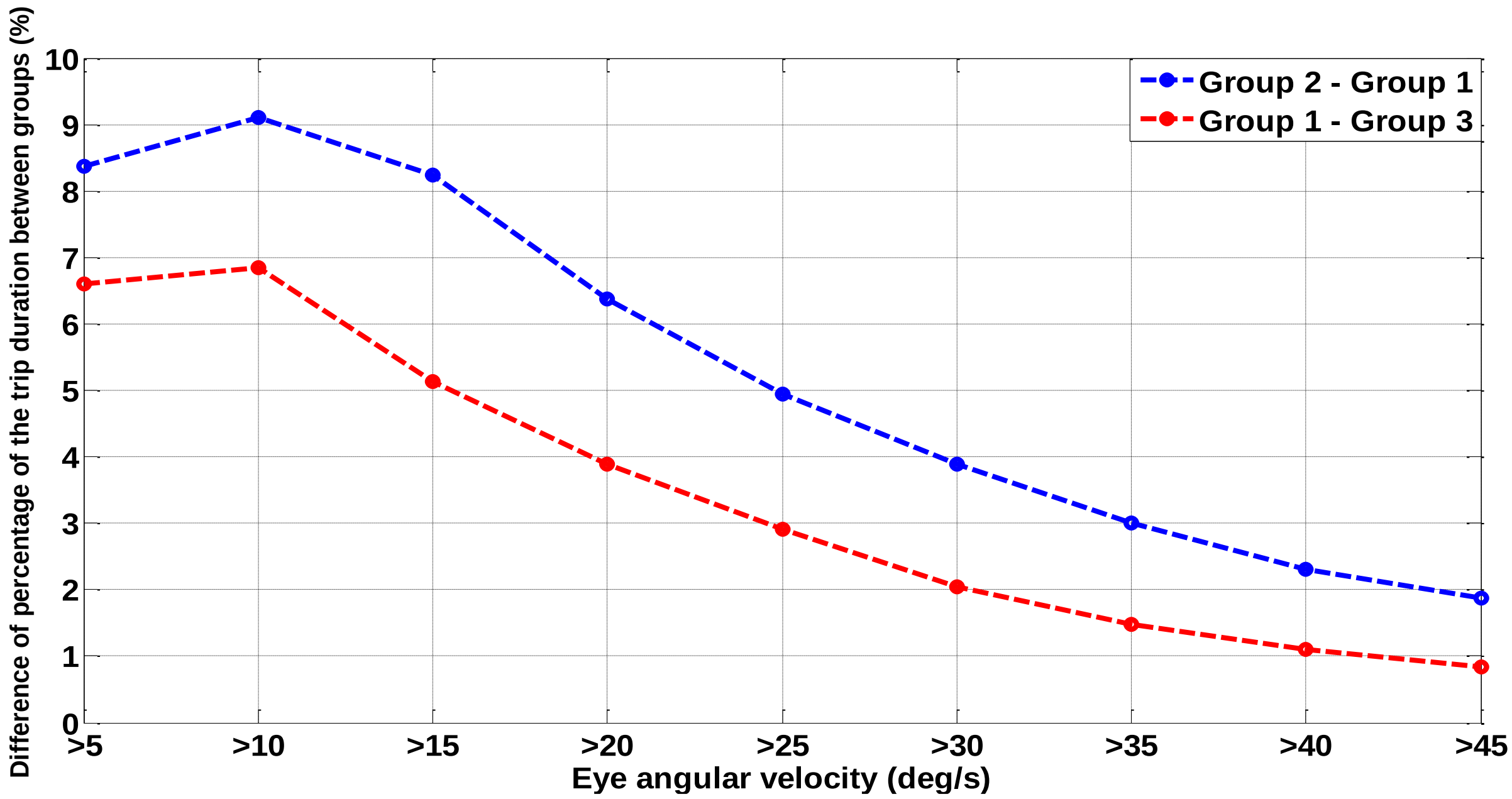

**Figure 8.** The difference in the percentage of time spent on the journey between the pairs of group 2 from group 1 and group 1 from group 3 where the angular velocity of the pupil movement is greater than the 9th angular velocity threshold.

Visual fixation analysis further supported these findings. Figure 10 and the surface plot in Figure 11 demonstrate an inverse relationship with angular velocity: Group 2 exhibited lower fixation durations compared to Group 1, while Group 3 showed slightly higher fixation durations than Group 1. The maximum differentiation between Groups 1 and 2 occurred at a 6-second fixation time within a 7 cm radius (Figure 13), whereas the maximum differentiation between Groups 1 and 3 was observed at a 4-second fixation time within an 8 cm radius (Figure 14).

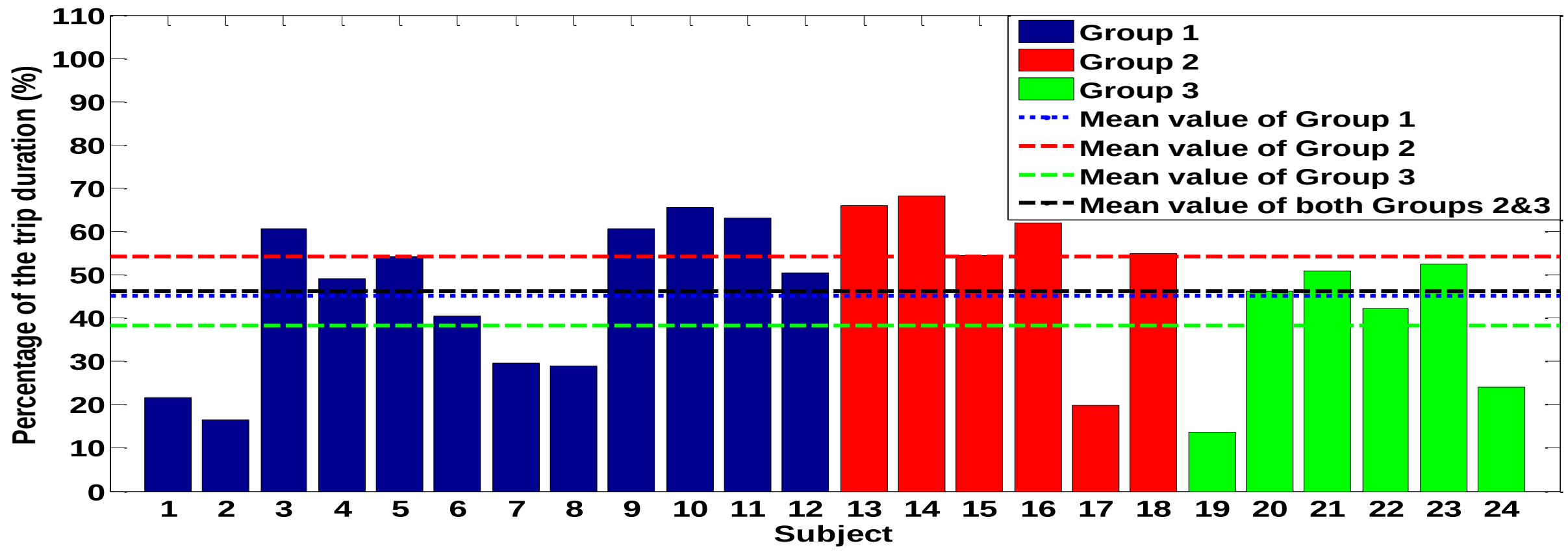

**Figure 9.** Percentage of time during the journey that the angular velocity of the pupil movement is greater than deg/s

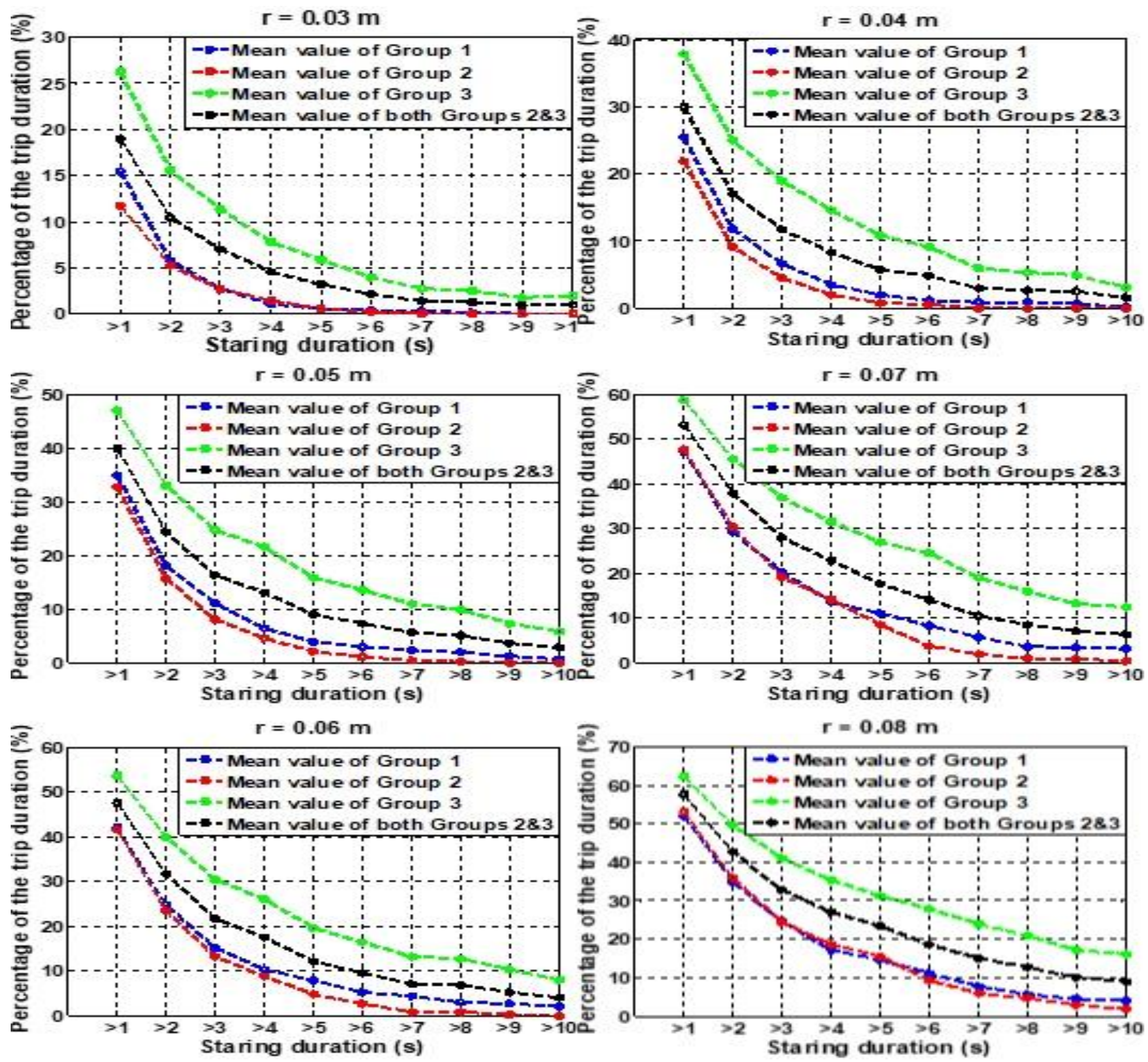

**Figure 10.** Angina Index: Percentage of duration of glare during travel for different glare time thresholds and glare areas with a radius of 3 to 8 cm for individuals in three groups

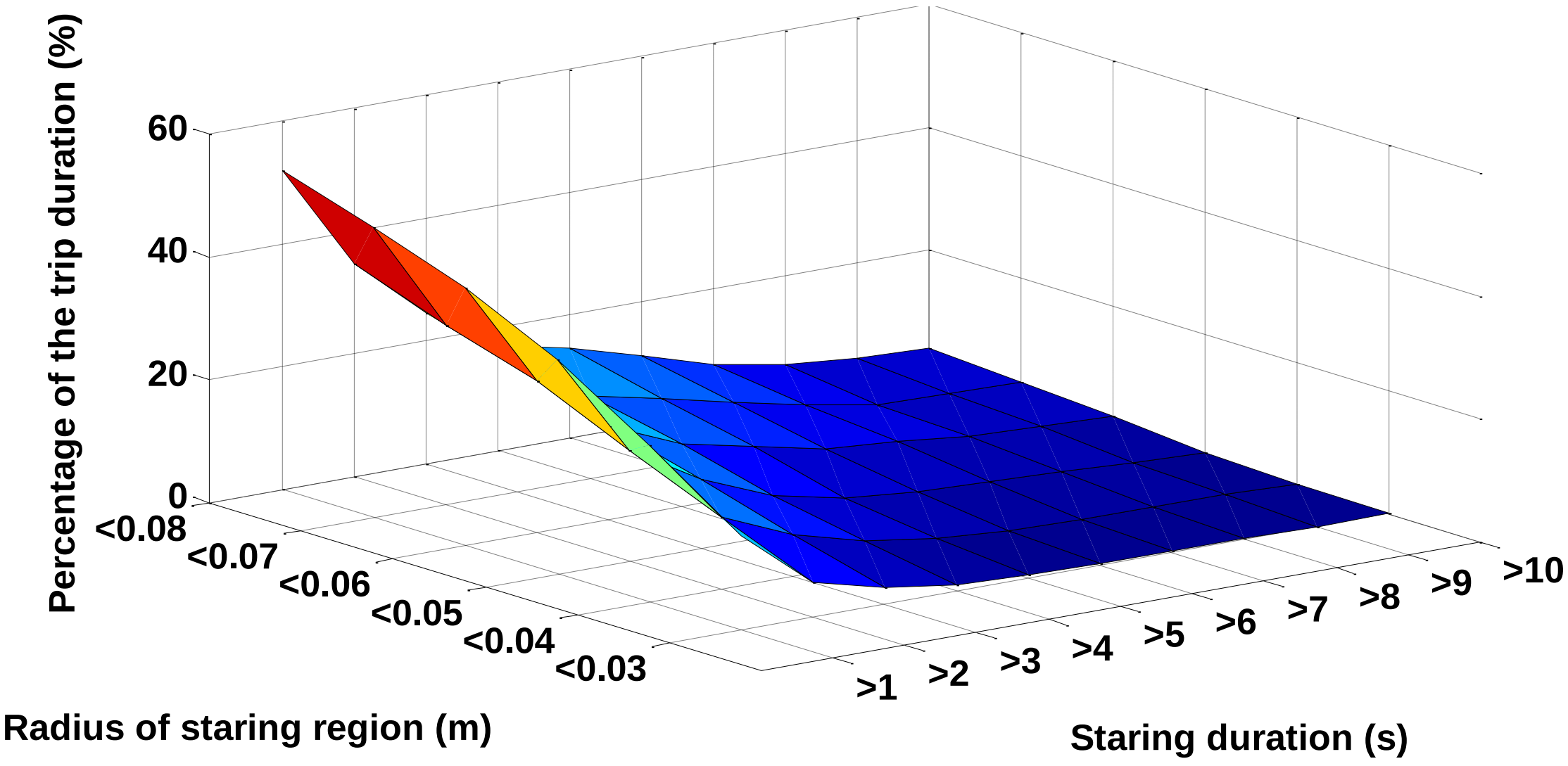

**Figure 11 .** Average percentage index of glare duration during travel for 10 glare time thresholds and 6 glare radius thresholds for individuals in the first group

To comprehensively identify the optimal parameters for this classification, the precise percentage differences in fixation durations between Group 1 and Group 3 were mapped across all evaluated time and radius thresholds, as visualized in the surface plot in Figure 12.

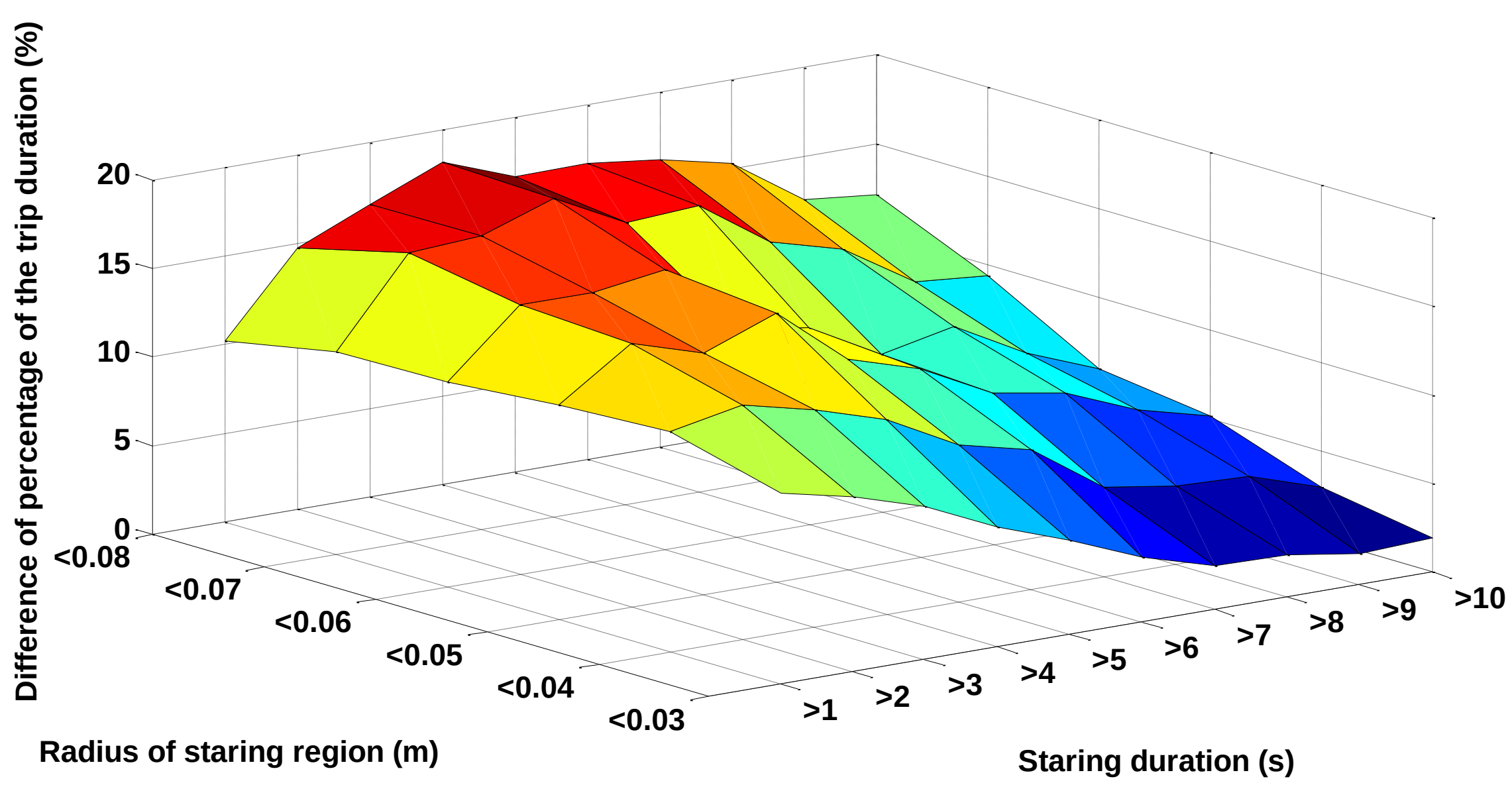

**Figure12** .Percentage difference in stare duration of group 3 from group 1 for 10 stare time thresholds and 6 stare radius thresholds

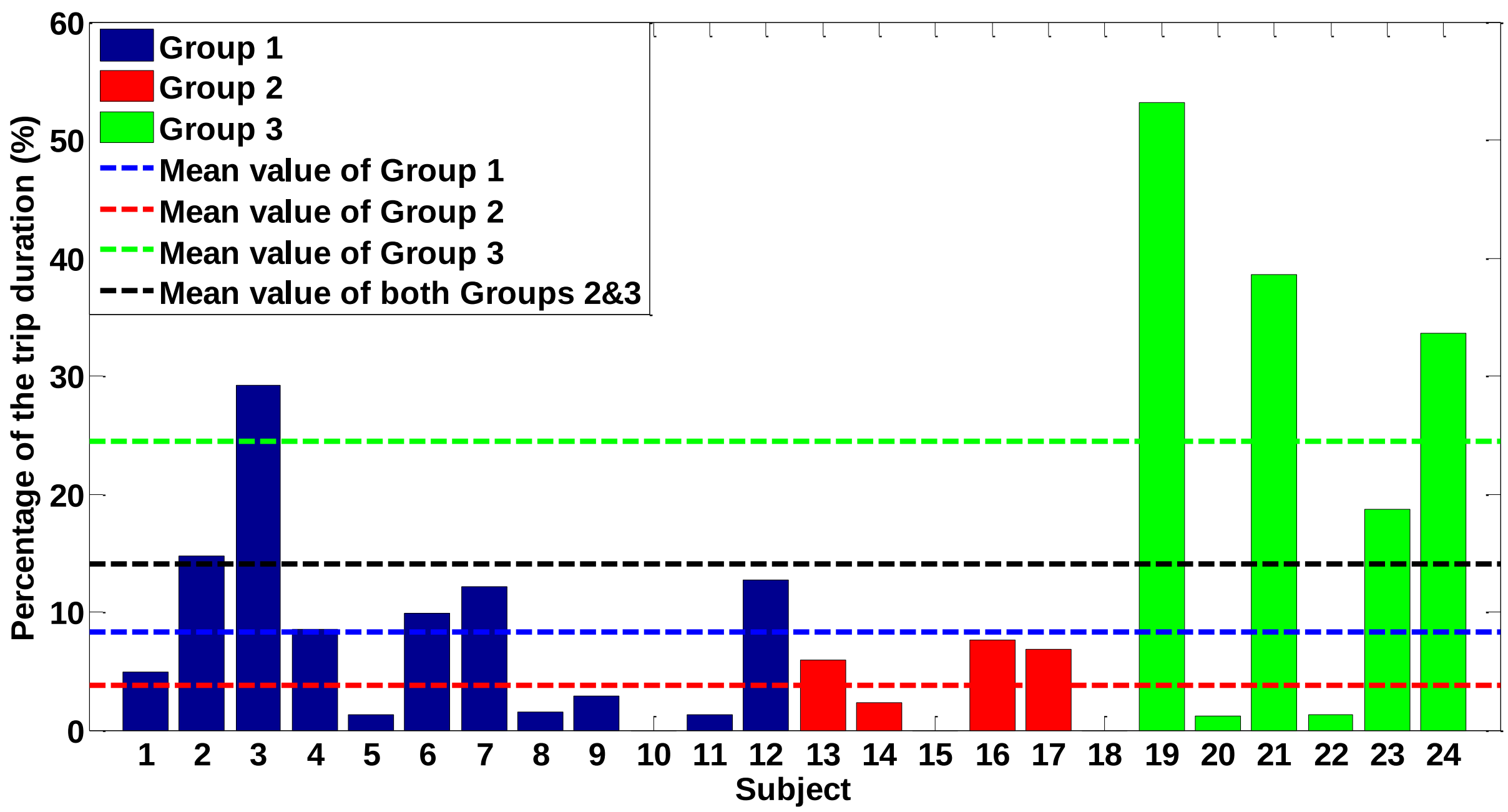

**Figure13** .Percentage of time during the trip that the eyes are dazzled for a glare time threshold of 6 seconds and a glare radius threshold of 7 cm

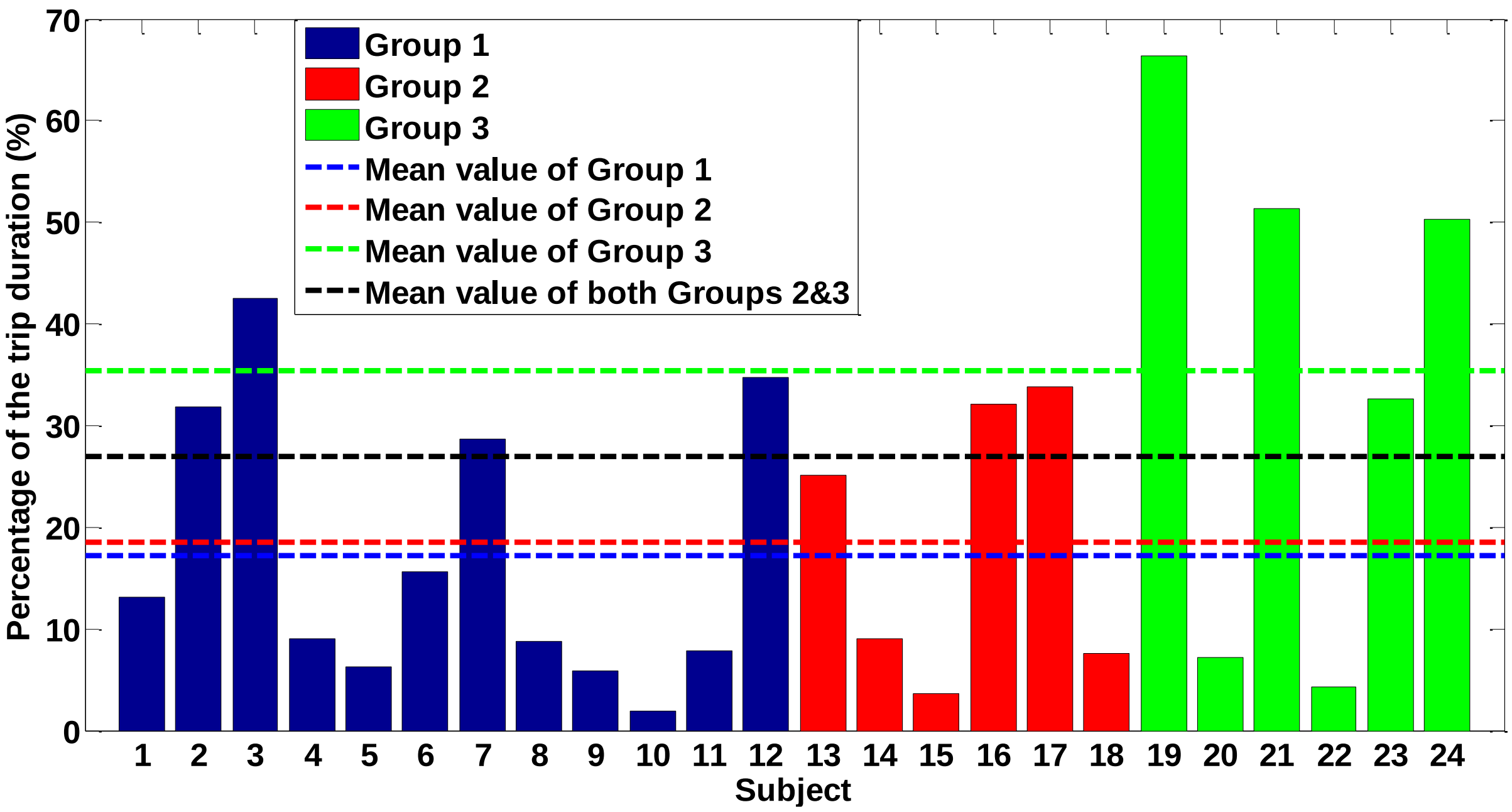

**Figure14** .Percentage of time during the trip that the eyes are dazzled for a glare time threshold of 4 seconds and a glare radius threshold of 8 cm

### 3.3.Classification Performance and Model Selection

The performance of the KNN classifier using various combinations of the four features and different values of K is summarized in Table 5. Models achieving an accuracy exceeding 70% are highlighted. For distinguishing between the control group (Group 1) and recovering patients (Groups 2 and 3 combined), the model trained solely on Feature 4 with K=2 achieved a notable accuracy of 75.4%. Feature 4 (left elbow displacement) proved to be the most critical independent variable for this classification task. When classifying Group 1 against Group 2, the model utilizing Feature 4 with K=3 achieved the highest accuracy of 87.8%. Conversely, distinguishing Group 1 from Group 3 proved more challenging, with the best model (using Features 2, 3, and 4 with K=3) reaching an accuracy of 74.6%. This lower accuracy suggests that the behavioral patterns of long-term treatment patients (Group 3) are more analogous to those of healthy drivers than those of early-stage patients (Group 2).

**Table5** .Performance of models with different features and different number of neighbors

| Different combinations of features | | | | | | | | | | | | | | | | |
|---|---|---|---|---|---|---|---|---|---|---|---|---|---|---|---|---|
| **1-2-3-4** | **2-3-4** | **1-3-4** | **1-2-4** | **1-2-3** | **3-4** | **2-4** | **2-3** | **1-4** | **1-3** | **1-2** | **4** | **3** | **2** | **1** | | |
| 61.0 | 58.9 | 54.6 | 61.9 | 49.9 | 60.4 | 67.0 | 46.3 | 63.1 | 46.0 | 47.0 | 75.4 | 39.0 | 42.6 | 10 | **1,2+3** | **k=2** |
| 53.6 | 53.3 | 51.0 | 55.9 | 48.9 | 52.4 | 59.3 | 45.9 | 55.6 | 43.3 | 45.0 | 57.3 | 31.7 | 49.6 | 35.3 | **1,2,3** | |
| 64.2 | 68.8 | 65.4 | 71.6 | 61.6 | 66.4 | 78.6 | 63.6 | 76.8 | 55.8 | 61.0 | 85.4 | 52.2 | 58.6 | 54.8 | **1,2** | |
| 60.0 | 62.2 | 62.2 | 60.8 | 61.6 | 62.4 | 63.8 | 62.2 | 63.0 | 59.4 | 59.6 | 65.6 | 62.6 | 62.8 | 51.2 | **1,3** | |
| 62.9 | 64.3 | 66.0 | 65.6 | 53.3 | 69.1 | 71.4 | 48.4 | 61.7 | 45.7 | 45.3 | 71.2 | 37.1 | 56.1 | 39.7 | **1,2+3** | **k=3** |
| 61.1 | 61.0 | 54.3 | 56.4 | 45.4 | 61.7 | 66.4 | 50.1 | 55.3 | 41.0 | 48.1 | 59.3 | 40.9 | 47.0 | 35.4 | **1,2,3** | |
| 72.6 | 73.2 | 71.6 | 84.2 | 60.4 | 75.6 | 86.0 | 58.2 | 81.0 | 56.8 | 60.8 | 87.8 | 45.2 | 61.6 | 56.8 | **1,2** | |
| 65.4 | 74.6 | 62.2 | 60.8 | 67.8 | 72.8 | 73.0 | 66.6 | 62.4 | 56.6 | 63.4 | 63.0 | 64.4 | 65.6 | 49.6 | **1,3** | |
| 58.4 | 60.0 | 55.4 | 59.4 | 48.1 | 66.6 | 67.9 | 43.7 | 61.0 | 47.1 | 45.6 | 74.9 | 39.3 | 42.4 | 40.1 | **1,2+3** | k=4 |
| 56.6 | 61.0 | 53.9 | 54.9 | 44.4 | 60.3 | 68.9 | 43.4 | 55.9 | 40.3 | 45.0 | 57.9 | 38.6 | 42.7 | 36.4 | **1,2,3** | |
| 61.8 | 68.6 | 66.8 | 71.2 | 58.4 | 67.8 | 80.8 | 55.8 | 75.0 | 63.2 | 59.2 | 79.8 | 54.2 | 57.2 | 59.8 | **1,2** | |
| 64.0 | 74.4 | 64.8 | 62.6 | 61.2 | 65.8 | 68.8 | 66.0 | 62.0 | 59.2 | 61.2 | 63.6 | 63.2 | 60.2 | 54.4 | **1,3** | |
| 64.9 | 67.0 | 57.3 | 57.1 | 46.7 | 71.3 | 67.7 | 36.3 | 57.0 | 47.7 | 43.0 | 74.6 | 41.3 | 38.7 | 34.6 | **1,2+3** | k=5 |
| 51.1 | 59.4 | 53.7 | 53.4 | 44.3 | 61.0 | 68.6 | 44.7 | 50.4 | 41.1 | 41.9 | 62.3 | 42.9 | 41.1 | 33.0 | **1,2,3** | |
| 72.6 | 72.6 | 70.2 | 69.4 | 55.6 | 71.0 | 83.8 | 47.8 | 74.6 | 57.2 | 55.0 | 85.2 | 56.2 | 52.6 | 57.4 | **1,2** | |
| 74.0 | 68.2 | 66.6 | 67.6 | 70.6 | 66.0 | 70.4 | 68.4 | 62.0 | 64.8 | 60.2 | 63.8 | 70.2 | 68.0 | 54.6 | **1,3** | |

The derived performance metrics are presented in Table 6. The model demonstrated an exceptional Specificity (True Negative Rate) of 0.99 and a Precision of 0.96, indicating an extremely low Type I error (False Positive Rate = 0.01). The overall model accuracy stood at 0.90. Finally, the Receiver Operating Characteristic (ROC) curve (Figure 15) illustrates the model's design point positioned favorably in the upper-left quadrant, confirming its high diagnostic capability and robustness.

**Table 6.**The values of the TPR sensitivity criteria, TNR detection, PPV accuracy, FPR false positive group detection rate, and ACC accuracy of the designed classification algorithm

| Performance metrics | Value |
|---|---|
| **Sensitivity** | 0.73 |
| **Specificity** | 0.99 |
| **Precision** | 0.96 |
| **False Positive Rate** | 0.01 |
| **Accuracy** | 0.90 |

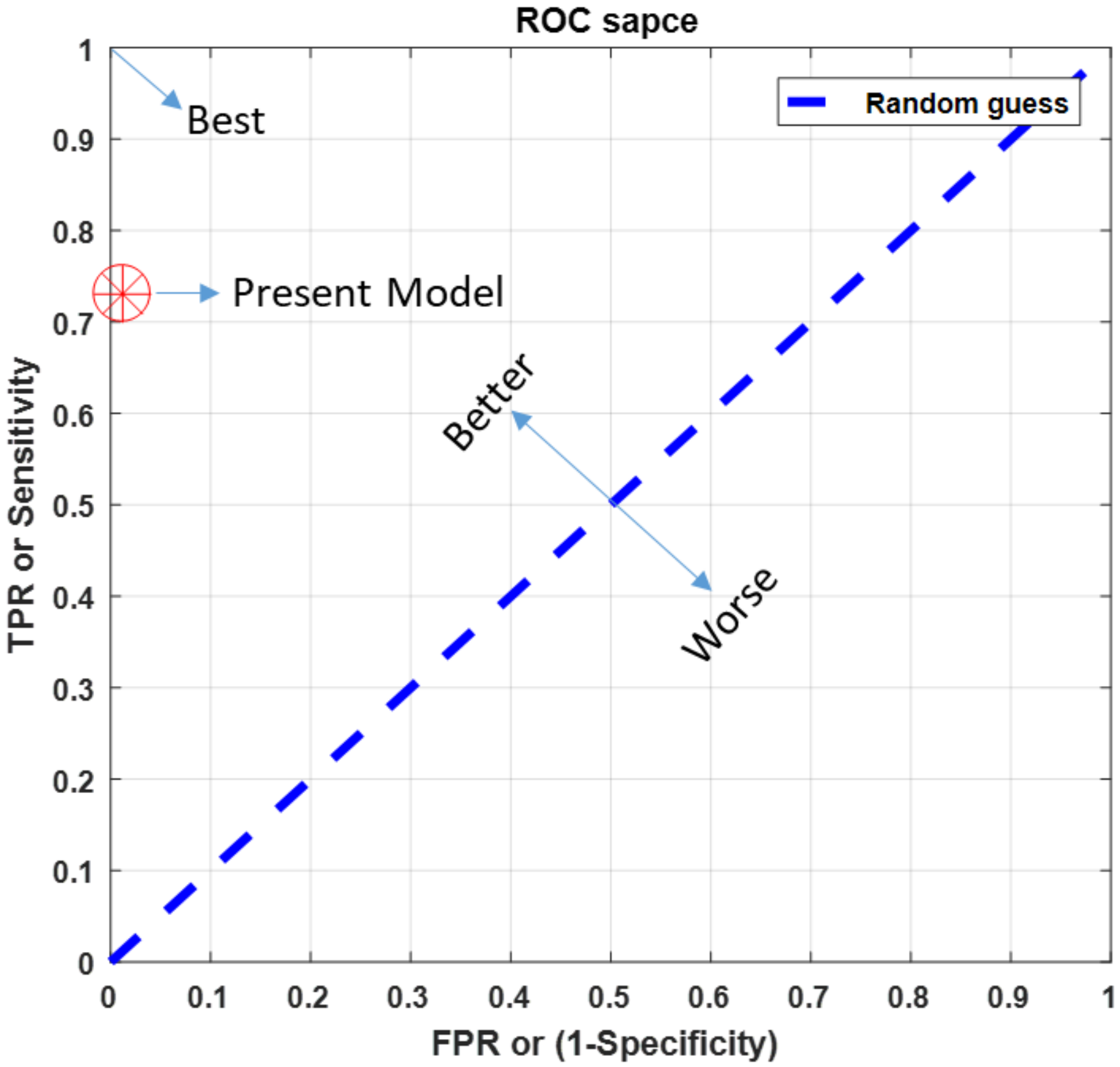

**Figure15** .Receiver performance characteristic point for the designed classification algorithm

## 3.4.Optimal Model Evaluation

Based on the comparative analysis, the KNN model utilizing Feature 4 and K=3 for classifying Groups 1 and 2 was selected as the optimal architecture. The comprehensive performance of this model over 100 validation iterations is detailed in the confusion matrix (Table 7).

**Table 7.** Confusion matrix for classifying individuals into groups 1 and 2

| Actual Class | Predicted: Normal (Negative) | Predicted: Group 2 Passenger (Positive) |
|---|---|---|
| **Normal (Negative)** | TN = 329 | FP = 4 |
| **Group 2 Passenger (Positive)** | FN = 45 | TP = 122 |

## 4. Discussion

The present study is among the few investigations that specifically focused not on individuals actively using amphetamine and their ability to drive safely, but on those undergoing the recovery process following amphetamine dependence.

Using an eye-tracking system and a Kinect sensor, the oculomotor and biomechanical movement of individuals in recovery was assessed in a driving simulator to develop a KNN model capable of distinguishing these individuals from healthy controls. Such an approach enables active monitoring of individuals who are at the highest risk of relapse, thereby facilitating timely interventions when necessary.

Previous studies have consistently demonstrated that psychostimulant use impairs driving performance and substantially increases the risk of traffic crashes, severe injuries, and fatalities [3]. To address this issue, various countermeasures have been implemented, including random roadside drug testing by law enforcement agencies, the use of drug detection devices, and the enforcement of financial and legal penalties for offenders [23]. However, these strategies are largely reactive and can be categorized as passive approaches to traffic safety.

Given that only a limited number of previous studies have included individuals undergoing substance withdrawal, the present study demonstrates that analyzing the eye and body movements of individuals in recovery may provide a proactive approach to improving road safety. Integrating these behavioral indicators into ADAS could enable earlier detection of driving impairments and help prevent traffic crashes. Furthermore, when appropriate and compliant with legal and ethical frameworks, these indicators could support monitoring programs implemented by relevant authorities, thereby enhancing traffic safety for this high-risk population. Here, we called it a proactive driving safety. Using eye tracker, Kinect and other sensors are well documented as driving monitoring systems [28, 29].

One of the most notable findings was that the displacement of the left hand (elbow) in individuals undergoing treatment was more than twice that observed in healthy participants, making it the most influential feature in the KNN classification model. Psychomotor stimulants are a class of drugs that act on the central nervous system (CNS) by increasing alertness, enhancing mood, and producing a sense of well-being [30]. The excessive left-hand movement observed in the recovery group may reflect elevated stress, restlessness, or impaired fine motor control when performing cognitively demanding driving tasks, such as maneuvering to avoid obstacles. These findings suggest that subtle alterations in upper-limb motor behavior may serve as objective behavioral markers of residual cognitive and motor impairments during the recovery process [31].

Besides physical movements, Participants in Group 2 (treatment duration <2 months) exhibited substantially faster eye movements and shorter fixation durations. In contrast, participants in Group 3 (treatment duration >2 months) demonstrated slower eye movements and longer fixation durations, exceeding those observed even in the healthy control group.

Change in eye-movement behaviour of amphetamine-addicted people and its role in driving is documented very well [28]. Of course, these visual effects were not always noticeable or at least did not always have a negative impact on driving performance [32].

In the current study for Group 2 (treatment duration <2 months), the pattern of faster eye movements and shorter fixation durations may be interpreted within the framework of hypervigilance, a phenomenon frequently observed during the early stages of substance withdrawal [33]. Individuals in this phase tend to continuously scan their surroundings with rapid saccadic eye movements while exhibiting relatively short fixation durations. In contrast, the eye movement pattern observed in Group 3 (treatment duration >2 months) may reflect the pharmacological effects of long-term OT maintenance therapy rather than the hyperarousal state characteristic of early withdrawal. As a CNS depressant, prolonged OT administration may reduce arousal levels and contribute to psychomotor retardation, resulting in slower saccadic eye movements and longer fixation durations [34]. This pattern may indicate reduced cognitive alertness and slower visual information processing, causing participants to maintain their gaze on individual objects for longer periods before shifting attention. Although these prolonged fixations may superficially resemble enhanced visual attention, they are more likely to reflect decreased processing speed and reduced psychomotor responsiveness associated with the sedative effects of opioid maintenance therapy.

The present study has several limitations that should be acknowledged. First, the sample size was relatively small, comprising only 24 participants. Consequently, the findings should be interpreted with caution, and their generalizability remains limited .Second, the study employed a low-fidelity driving simulator, therefore, the ecological validity of the findings may be limited.

## 5. Conclusion and Practical Implications

The behavioral features identified in the present study have the potential to serve as non-intrusive inputs for ADAS. If the vehicle detects behavioral patterns indicative of reduced attention, heightened restlessness, or impaired psychomotor performance resulting from clinical conditions or pharmacological treatment, the ADAS could adapt its assistance strategy accordingly. For example, the system could provide tailored warning modalities or, in partially automated driving scenarios, allocate a longer and more individualized take-over request interval, allowing drivers additional time to safely regain control of the vehicle. Also, these behavioral indicators could also support broader traffic safety initiatives. Subject to appropriate legal authorization, ethical oversight, and privacy safeguards, relevant risk-related information could be securely shared with external traffic safety authorities, such as highway police. Such integration could facilitate proactive monitoring of high-risk drivers undergoing substance-use treatment and enable timely assistance or preventive interventions, when necessary, thereby reducing the likelihood of traffic crashes while respecting individual rights and data protection requirements.

**Conflict of interest:** None

**Funding:** This research did not receive any specific grant from funding agencies in the public, commercial, or not-for-profit sectors.

**Declaration of generative AI and AI-assisted technologies:** In the final preparation of this manuscript, Chat GPT-5 is used to check the grammar and increase the readability of the content. After that, all the content is re-checked by the corresponding author to ensure about its intended meaning.